\documentclass[%
 reprint,
 amsmath,amssymb,
 aps,
pra,
]{revtex4-1}
\usepackage{listings}

\usepackage{graphicx}
\usepackage{dcolumn}
\usepackage{bm}
\usepackage{natbib}
\usepackage[colorlinks,citecolor=red,linkcolor=blue,urlcolor=blue]{hyperref}
\usepackage{caption}
\usepackage{hhline}
\usepackage{float}
\usepackage{mhchem}
\usepackage{calrsfs}
	\DeclareMathAlphabet{\pazocal}{OMS}{zplm}{m}{n}
\usepackage{physics}
\usepackage{appendix}
\usepackage[export]{adjustbox}

\newcommand{\bk}{\bold{k}}

\usepackage{fancyhdr}
\begin{document}


\title{Twisted superfluid and supersolid phases of triplons in bilayer honeycomb magnets}
\author{Dhiman Bhowmick$^1$ \href{https://orcid.org/0000-0001-7057-1608}{\includegraphics[scale=0.12]{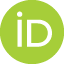}}}
\author{Abhinava Chatterjee$^{1,2,*}$}
\author{Prasanta K. Panigrahi$^2$}
\author{Pinaki Sengupta$^1$\href{https://orcid.org/0000-0003-3312-2760}{\includegraphics[scale=0.12]{orcid.png}}}%
\affiliation{%
 $^1$School of Physical and Mathematical Sciences, Nanyang Technological University, 21 Nanyang Link, Singapore 637371, Singapore \\
}%
\affiliation{%
 $^2$Department of Physical Sciences, Indian Institute of Science Education and Research Kolkata, Mohanpur - 741246, India \\
}%
\thanks{Present address: Department of Physics, Yale University, New Haven, Connecticut 06520, USA}
\date{\today}

\begin{abstract}
We demonstrate that low-lying triplon excitations in a bilayer Heisenberg antiferromagnet provide a promising avenue to realise magnetic analogs of twisted superfluid and supersolid phases that were recently reported for two-component ultracold atomic condensate in an optical lattice. 
Using a cluster Gutzwiller mean field theory, we establish that Dzyaloshinskii-Moriya interactions (DMI), that are common in many quantum magnets, stabilize these phases in magnetic system, in contrast to pair hopping process that is necessary for ultracold atoms.
The critical value of DMI for transition to the twisted superfluid and twisted supersolid phases depends on the strength of the (frustrated) interlayer interactions that can be tuned by applying external pressure on and / or shearing force between the layers. Furthermore, we show that the strength of DMI can be controllably varied by coupling to tailored circularly polarized light. Our results provide crucial guidance for the experimental search of twisted superfluid and supersolid phases of triplons in real quantum magnets.
\end{abstract}

\pacs{Valid PACS appear here}
\maketitle


\section{\label{sec1}Introduction}
The observation of twisted, multi-orbital superfluid in  binary mixtures of ultracold $^{87}$Rb atoms in two different hyperfine states on a honeycomb optical lattice has attracted heightened interest in this novel quantum state of matter\,\cite{Rb}. 
The twisted superfluid (or twisted supersolid) state is characterised by a complex order parameter -- the phase of the local superfluid order parameter at each site changes continuously forming a``twisting pattern", thus breaking time reversal symmetry spontaneously. Interestingly, complex order parameters have experimentally been shown to be associated with other novel strongly correlated phases such as the superconducting states of \ce{Sr_2RuO_4}\,\cite{TRS_SC1,TRS_SC3} and \ce{UPt_3}\,\cite{TRS_SC2} and the pseudo-gap state in the cuprate high-Tc superconductor, B-2212\,\cite{TRS_SC4, TRS_SC5}. A detailed understanding of the twisted superfluid state can help gain insight into these states as well. Subsequent theoretical studies have shown that the extended Bose Hubbard model with an additional pair hopping term can stabilize a twisted superfluid (TSF) ground state over a finite range of parameters\,\cite{TSF}. 

Quantum magnets have long served as a versatile platform for realizing magnonic analogs of complex bosonic phases, often under less extreme conditions. For example, temperature needed for Bose Einstein condensation (BEC) of magnons varies from a few Kelvins in many quantum magnets\,\cite{BEC1,BEC2,BEC3} to room temperature in Yttrium Iron Garnet (YIG) thin films\,\cite{YIG_BEC1,YIG_BEC2}, in contrast to nano-Kelvin temperature scales required for BEC in ultracold atoms\,\cite{BEC_ATOM_1,BEC_ATOM_2}.
In this work, we show that twisted superfluid and twisted supersolid phases of magnons are realized in a bilayer honeycomb Heisenberg model. Interestingly, pair hopping process of magnons is not essential for stabilizing these phases\,\cite{TSF}, in contrast to ultracold atomic systems. Instead, a next nearest neighbor (NNN) Dzyaloshinskii-Moriya interaction\,(DMI) -- which is present in many quantum magnets -- is sufficient to yield field induced twisted superfluid (TSF) and twisted supersolid (TSS) phases over wide ranges of parameters.

Our paper is structured as follows. In Sec.\,\ref{secII}, we introduce the microscopic spin Hamiltonian and describe the physics in the non-interacting limit by deriving the tight-binding triplon Hamiltonian and triplon band structure.
The cluster Gutzwiller mean field theory\,(CGMFT) is introduced in Sec.\,\ref{secIII} which is used to calculate order-parameters in interacting limit. 
In Sec.\,\ref{secIV} we present the results of our study in the form of the order parameters and phase diagram as a function of magnetic field, DMI and Heisenberg interactions. 
In Sec.\ref{secV}, we propose the possible materials in search of TSS and TSF phases.
Finally, in Sec.\,\ref{secV}, we show that circularly polarized light can be used to induce DMI greater than the cutoff DMI required to realize TSS (TSF), whereas presence of frustration among interlayer and intralayer interactions is helpful to lower the cutoff value of DMI (see also Appendix.\,\ref{appendixC}).
The principal findings are summarized in section Sec.\,\ref{secVI}.

\begin{figure}[H]
	\centering
		\includegraphics[width=0.45\textwidth]{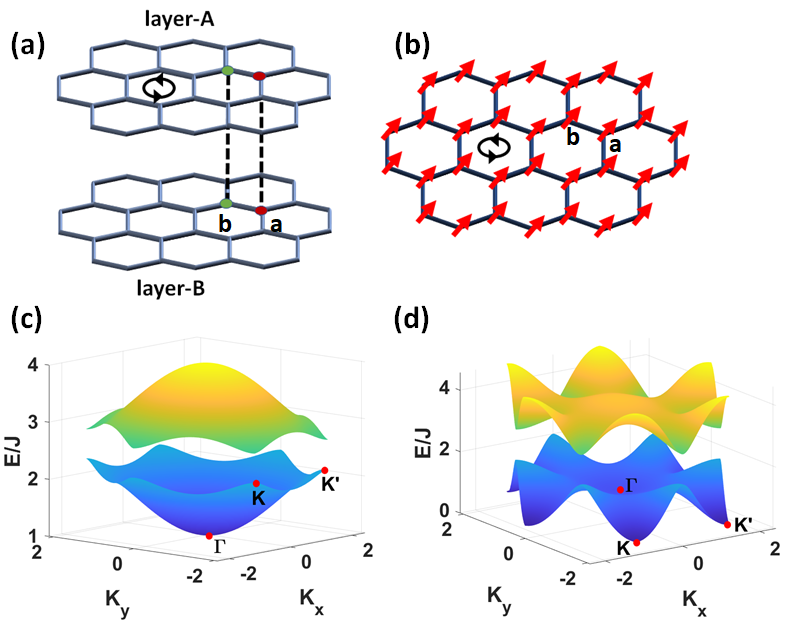}
	\caption{\,(Color online)\,(a) The spins on each lattice site interact via a strong interlayer anti-ferromagnetic coupling, resulting in a ground state of a honeycomb lattice of dimers on each interlayer nearest-neighbor bonds (shown by dashed-black line). (b) The ferromagnetic honeycomb lattice. (c) Triplon band structure at $D=0.1J$, (d) Triplon band structure at $D=0.8J$. The other parameters for the band structure are $J_\perp=10J$, $B_z=0.0$, $J_z=0.0$.}
	\label{Material}
\end{figure}

\section{\label{secII}The bilayer honeycomb magnet and effective triplon model}

We start with a $S=1/2$ Heisenberg antiferromagnet on a bilayer honeycomb lattice with out of plane exchange anisotropy and Dzyaloshinkii-Moriya interaction\,(DMI), schematically shown in Fig.\,\ref{Material}(a) and described by the Hamiltonian,
\begin{align}
\pazocal{H}=&J_\perp\sum_{\substack{i,\,m\in A\\ n\in B}}\bold{S}_{i,m}\cdot\bold{S}_{i,n}-B_z\sum_{i,m} S^z_{i,m}\nonumber\\
&+\sum_{\left\langle i,j\right\rangle,m}\left[J(S^x_{i,m}S^x_{j,m}+S^y_{i,m}S^y_{j,m})+J_z S^z_{i,m} S^z_{j,m}\right] \nonumber \\
&+ D\sum_{\left\langle\left\langle i,j\right\rangle\right\rangle,m} \nu_{ij} \hat{z}\cdot (\bold{S}_{i,m}\times \bold{S}_{j,m}).
\label{eq::SpinHamiltonian}
\end{align}
$\bold{S}_{i,m}$ denotes the spin operator at $i$-th interlayer bond at layer $m$\,($m\in \{A,B\}$). $\langle\ldots \rangle$ and $\langle\langle\ldots\rangle\rangle$ denote the nearest-neighbor (NN) and next-nearest-neighbor (NNN) inter-dimer bonds respectively.
$J_\perp(>0)$ is the strength of (isotropic) interlayer Heisenberg interaction, while
$J_z$\,($>0$) and $J$\,($>0$) denote the Ising and XX type intralayer NN Heisenberg exchange interactions respectively. $D$ is the Dzyaloshinskii–Moriya interaction\,(DMI) which is constrained by the symmetry of the lattice to intralayer NNN bonds; $\nu_{ij}=+1$, if $i\to j$ forms part of a counterclockwise closed loop connecting the NNN sites in a hexagonal plaquette in each layer\,(see counterclockwise circular arrows in Fig.\,\ref{Material}(a)-(b)) and $\nu_{ij}=-1$ otherwise. 
Finally $B_zS_{i,m}^z$ describes a Zeeman coupling between a local spin-moment and an external longitudinal magnetic field. 
It is noticeable that the change in sign of $J$ and $D$ do not alter the magnetic ground state as well as excitations above the ground state, whereas $J_z$ and $J_\perp$ are strictly positive in this study.

For $J_{\perp}\gg |J|$, the ground state of the system is a product of the singlet dimers on each interlayer NN bond. 
In this limit, the lowest excitations of the system are triplons, which are localized $S=1$ quasipartices.
An out-of-plane magnetic field lowers the energy of the $S_z=+1$ triplons and at a critical magnetic field, it crosses the energy of the singlet state, populating the ground state with a finite density of triplons. 
The other triplon branches\,($S^z=0$ and $S^z=-1$) are separated by a large energy gap. At low temperatures, one can restrict the local Hilbert space of the dimers to the singlet and $S^z=+1$ triplon. 
By  treating the triplons as bosonic quasiparticles, one can formulate a description of the low energy physics of the system in terms of hard core bosons which is known as bond-operator formalism\,(see also Appendix.\ref{appendixA}).
In this formalism, the zero field ground state made of singlets on each interlayer bonds is considered an empty lattice with number of triplons $n_i =0\; \forall i$.
At the critical field, a finite density of triplons is generated which increases with increasing field.  
The inter-dimer exchange interactions induce an effective hopping of the triplons. 
This delocalization induces a BEC of $S_z=+1$ triplons in the ground state. 
In the spin language, this corresponds to an canted antiferromagnetic order with a spontaneously broken U(1) symmetry. 
Considering singlets as a vacuum state in the system and triplon\,($S_z=+1$) as a hard-core bosonic quasi-particle excitations in vacuum of singlets, we can use  the bond operator formalism to express the bi-layer spin Hamiltonian Eq.\,\ref{eq::SpinHamiltonian} as an effective triplon Hamiltonian on a single-layer honeycomb lattice\,\cite{ZapfBatista, bond_operator, bond_operator2}\,(see Appendix.\,\ref{appendixA}),
\begin{align}
\pazocal{H}=&\frac{J}{2}\sum_{\left\langle ij\right\rangle} \left[\hat{t}^\dagger_i\hat{t}_j+\text{H.c.}\right]
+\frac{\mathfrak{i}D}{2}\sum_{\left\langle\left\langle ij\right\rangle\right\rangle} \nu_{ij} \left[\hat{t}^\dagger_i\hat{t}_j-\text{H.c.}\right]\nonumber\\
&+\left(\frac{J_\perp}{4}-B_z\right)\sum_i \hat{t}^\dagger_i \hat{t}_i
+\frac{J_z}{2}\sum_{\left\langle ij\right\rangle} \hat{n}_i \hat{n}_j
\label{Hamiltonian}
\end{align}
 where, $\hat{t}_i$ is the triplon annihilation operator, $\hat{n}_i$ is the triplon number operator and index-$i$ represents site index of effective single-layer honeycomb lattice\,(which is equivalent to interlayer bond-index of bi-layer honeycomb lattice). 
The first two terms represent hopping of triplons between NN and NNN neighbor dimers respectively, the third term is an on-site potential (effectively a chemical potential) and the last term describes the effects of NN interaction between triplons.
It is noticeable that the NNN hopping has a complex weight which renders the Hamiltonian unsuitable for quantum Monte Carlo simulations.
When $J_z\approx 0$, Eq.\ref{Hamiltonian} reduces to a tight binding model of non-interacting triplons. The Bloch Hamiltonian in the momentum basis, in terms of the momentum space triplon operators, is determined via Fourier transformation as,
 \begin{equation}
 \pazocal{H}=\sum_{\bold{k}} \Psi_{\bold{k}}^\dagger  \left[g(\bold{k})\sigma_0+\bold{h}\cdot\boldsymbol{\sigma}\right]\Psi_{\bold{k}},
 \label{eq::Hkequation}
 \end{equation}
where, $\Psi_{\bold{k}}=(\hat{a}_{\bold{k}},\hat{b}_{\bold{k}})^T$.
$\hat{a}_{\bold{k}}$\,($\hat{b}_{\bold{k}}$) denotes the ${\bf k}$-space triplon annihilation operator on sublattice-a\,(b) as shown in Fig.\,\ref{Material}(a)-(b) and $\boldsymbol{\sigma}$ is the pseudo-vector of Pauli matrices and $\sigma_0$ is the two-dimensional identity matrix. The coefficients of the $\sigma$-matrices in the Bloch Hamiltonian are $g(\bold{k})=(J_\perp/4)-B_z$, $h_x(\bold{k})=(J/2)\sum_{i} \cos(\bold{k}\cdot\boldsymbol{\alpha}_i)$, $h_y(\bold{k})=(J/2)\sum_{i} \sin(\bold{k}\cdot\boldsymbol{\alpha}_i)$, $h_z(\bold{k})=D\sum_{i} \sin(\bold{k}\cdot\boldsymbol{\beta}_i)$, where $\boldsymbol{\alpha}_i$ and $\boldsymbol{\beta}_i$ are the NN and NNN vectors for each layer respectively. $\boldsymbol{\beta}_i$'s are chosen such that they form a   counter-clockwise triangular loop for sites in sublattice-a in a hexagonal plaquette and clockwise triangular loop for sites in sublattice-b. The energy eigenvalues are given by,
\begin{equation}
E^{\pm}(\bold{k})=g(\bold{k})\pm |h(\bold{k})|.
\end{equation}
The band dispersion is shown in Fig.\,\ref{Material}(c) and Fig.\,\ref{Material}(d) for two values of DMI. For ${\bf D} = 0$, the energy spectrum is identical to that of graphene, with a linear band crossing of the upper and lower bands at the Dirac points $K$ and $K'$.  A finite DMI breaks time reversal symmetry and opens a band-gap $6\sqrt{3}D$ at these points. 
The energy of the lower band at $\Gamma$ and $K$\,($K'$)-points are respectively given by (at $B_z=0$),
\begin{equation}
E_{\Gamma}=\frac{J_{\perp}}{4}-\frac{3|J|}{2},\,\,
E_{K}=\frac{J_\perp}{4}-\frac{3\sqrt{3}}{2}|D|.
\label{EnergyGap}
\end{equation}
In the absence of DMI, the energy minimum is located at the center of the Brillouin zone, the $\Gamma$ point. For a finite but small DMI, the band minimum remains at $\Gamma$-point\,(Fig.\,\ref{Material}(c)). 
Increasing DMI to $|D|>|J|/\sqrt{3}$ shifts the band minimum from $\Gamma$-point to two degenerate minima at the $K$ and $K'$\,(Fig.\,\ref{Material}(d)).
Thus with changing DMI, the ground state changes from a one component BEC (condensation momentum ${\bf k} = 0$) to a two component BEC (condensation momenta at {\bf k} = $K$ and $K'$). 
The transition happens at $|D|=|J|/\sqrt{3}$ independent of $J_\perp$ and $B_z$ for small $B_z$.

In presence of repulsive interaction between triplons\,(\,the last term in Eq.\,\ref{Hamiltonian}\,) co-existence of triplons at $K$ and $K'$ points costs no additional energy classically\,(\,see Appendix.\,\ref{appendixB}\,)\,, but the quantum-fluctuations around $\bold{K}$ and $\bold{K}^\prime$ points introduces an energy cost\,\cite{SymmetryBreaking1}.
Thus spontaneous breaking of valley-symmetry is energetically favoured and the quantum fluctuation will lead to a ground state with only single valley condensation which is known as ``quantum order by disorder" effect\,\cite{SymmetryBreaking1,SymmetryBreaking2,SymmetryBreaking3}.
Thus, in the limit of weak interaction, the super-fluid order parameter at lattice site $\bold{r}$ is either $|b|e^{i\bold{K}\cdot\bold{r}+\phi}$ or $|b|e^{i\bold{K}'\cdot\bold{r}+\phi}$ depending on the valley for the Bose-Einstein condensate, where $\phi$ is a global phase independent of position of lattice site.
This spontaneous breaking of valley symmetry transforms the superfluid order-parameter from a real to a complex value and the resultant BEC is known as twisted superfluid\,\cite{TSF}.
Based on this, the ground state on the honeycomb lattice is no longer a bipartite lattice in twisted superfluid phase, but become a lattice with six-sublattices as shown in the figure Fig.\,\ref{Cluster}.
Results based on CGMFT also supports this scenario as shown in section Sec.\,\ref{secIII}.


 In absence of interaction $J_z$, Eq.\,\ref{Hamiltonian} transforms into well-known bosonic topological Haldane model. 
 At finite temperatures due to non-zero magnetic excitations, this model is known to exhibit a finite thermal Hall effect due to presence of non-zero Berry-curvature of the bands.
 On this basis it is expected that Bose-Einstein condensate at $\Gamma$-point or $K$\,($K'$)-point would provide a non-zero Hall conductance even at zero temperature\,\cite{SymmetryBreaking1}.
 However, we find that the Berry-curvature at the condensation momenta -- $\Gamma$-point when $|D|<|J|/\sqrt{3}$ and $K$\,($K'$)-point when $|D|>|J|/\sqrt{3}$ -- is zero.
 Since the density of magnons is concentrated around the condensation momentum, and is vanishingly small away from it, the thermal Hall response of the Bose-Einstein condensate is vanishingly small in our model.

An analogous bosonic model can also be observed in terms of Matsubara-Matsuda bosons in magnetically ordered honeycomb ferromagnets\,(see Fig.\,\ref{Material}(b)\,) or antiferromagnets.
Thus the non-trivial phases like twisted-superfluid or twisted-supersolid are also expected to emerge in magnetically ordered systems.
However, in practice, the presence of anisotropies in magnetically ordered systems break the $U(1)$-symmetries destroying conservation of numbers of particles which in-turn preclude a long time superfluidity response in this systems\,\cite{ZapfBatista,Giamarchi_2008}.
Moreover due to presence of strong exchange interactions\,($J,\,J_z$ in Eq.\,\ref{Hamiltonian}) in general, the phase transitions are also difficult to study varying the external parameters in these systems\,\cite{Giamarchi_2008}. 
That is why the dimerized paramagnets are well suited to study superfluid phases and superfluid-Mott transitions.
Hence we focus on bilayer dimerized honeycomb paramagnetic system to study the twisted superfluid and twisted supersolid phases. 


\section{\label{secIII} Cluster-Gutzwiller Mean Field Theory (CGMFT) and Observables}

\begin{figure}[H]
	\centering
		\includegraphics[width=0.47\textwidth]{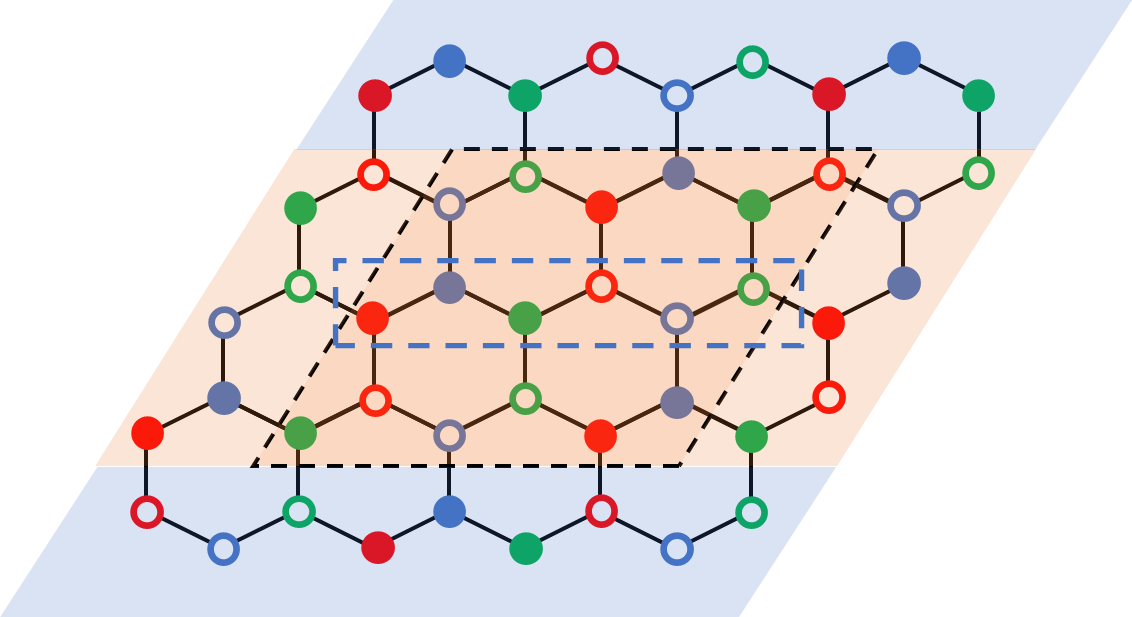}
	\caption{\,(Color online)\,The cluster construction for CGMFT. There are 18 sites in the cluster which is located within the dashed black box. A periodic boundary condition is applied along the horizontal direction and the mean-field boundary condition is applied along the vertical direction. The background of the cluster sites is denoted by a pink shade and the background of the mean-field sites are denoted by blue shade.}
	\label{Cluster}
\end{figure}

The cluster Gutzwiller mean field theory or CGMFT\,\cite{ClusterGutzwiller, TSF} -- equivalently, cluster mean field theory\,\cite{CMF,CMF1,CMF2,CMF4,CMF5,CMF6,CMF7}, self-consistent cluster mean field theory\,\cite{SelfConsistentCMF1}, multi-site mean field theory\,\cite{MultiSiteMF}, hierarchical mean field approach\,\cite{Hierarchical1,Hierarchical2}, composite boson mean field theory\,\cite{CompositeBoson} -- is a powerful technique to study superfluid phases in bosonic many body systems with complex hopping terms.
CGMFT improves over the conventional single-site mean field approach by taking into account the short range correlations present within a small lattice-cluster using exact-diagonalization.
Furthermore it is an alternative numerical method to study the quantum systems like we described in section Sec.\,\ref{secII}, where sign problems arises in quantum Monte-Carlo methods due to complex hopping terms or geometric frustration\,\cite{Frustration, Hierarchical}.
 Whereas conventional mean-field theories\,\cite{MeanField} and exact-diagonalization of small systems\,\cite{ED} fail to show existence of TSF and TSS phases, Density Matrix Renormalization Group for one dimensional systems\,\cite{DMRG} and CGMFT for higher dimensional systems\,\cite{TSF} are better alternatives  for search of these non-trivial phases.

We explore the ground state phases of the effective triplon Hamiltonian with CGMFT by decomposing the system into clusters\,(pink shaded region) and mean-field region\,(blue shaded region) as shown in figure Fig.\,\ref{Cluster}.
The effective mean-field Hamiltonian of the cluster is given as,
\begin{equation}
	\pazocal{H}_C^{\text{eff}}= \pazocal{H}_C + \pazocal{H}_{\delta C},
\end{equation}
where, $\pazocal{H}_C$ is the Hamiltonian as in equation Eq.(\,\ref{Hamiltonian}) within the cluster and $\pazocal{H}_{\delta C}$ is the Hamiltonian which takes into account the interactions among the boundary sites of the cluster and the mean-field region.
The form of the boundary Hamiltonian is given by,
\begin{align}
	\pazocal{H}_{\delta C}=
\frac{J}{2} &\sideset{}{'}\sum_{\left\langle i,j \right\rangle}
\left[ \hat{t}_i^\dagger \left\langle\hat{t}_j\right\rangle + \text{H.c.} \right]
+\frac{iD}{2} \sideset{}{'}\sum_{\left\langle\left\langle i,j \right\rangle\right\rangle}
\left[ \hat{t}_i^\dagger \left\langle\hat{t}_j\right\rangle - \text{H.c.} \right]
\nonumber\\
&+\frac{J_z}{2} \sideset{}{'}\sum_{\left\langle i,j \right\rangle}
 \hat{n}_i \left\langle\hat{n}_j\right\rangle,
\end{align}
where the primed summations are over the boundary site-$i$ connected to the mean-field site-$j$.
$\left\langle\hat{t}_j\right\rangle$ and $\left\langle\hat{n}_j\right\rangle$ are two mean-field parameters denoting the superfluid order parameter and occupation number of triplons at site-$j$ respectively.
We choose six inequivalent sites in each cluster (denoted by different patterns in figure  Fig.\,\ref{Cluster}) to give a total of 12 mean-field parameters. The ground state in the different parameter regimes are obtained by evaluating these mean field parameters self-consistently in the following manner,
\begin{itemize}
\item[(i)] Choose an initial set of mean field parameters $\left\lbrace\left\langle\hat{t}_j\right\rangle,\, \left\langle\hat{n}_j\right\rangle\right\rbrace$, $j=1,\dots, 6$ and then exactly diagonalize the effective Hamiltonian of the cluster $\pazocal{H}^{\text{eff}}_C$.
\item[(ii)] Calculate new mean field parameters $\left\langle\hat{t}^\prime_j\right\rangle$ and $\left\langle\hat{n}^\prime_j\right\rangle$ from the sites within blue-dashed rectangle in the figure Fig.\,\ref{Cluster} which reside within the cluster. Periodic boundary condition is chosen along horizontal direction to eliminate any boundary effect on the sites within the blue-dashed rectangle, so that the mean field parameters obtained from those sites are free from boundary effects.
\item[(iii)] The initial and final set of mean field parameters are compared using the tolerance 
\begin{equation}
\epsilon=\sum_j \left|\left\langle\hat{t}^\prime_j\right\rangle-\left\langle\hat{t}_j\right\rangle\right|+\sum_j \left|\left\langle\hat{n}^\prime_j\right\rangle-\left\langle\hat{n}_j\right\rangle\right|.
\end{equation}
If the tolerance $\epsilon$ is less than a certain cutoff then the obtained mean-field parameters correspond to the ground state of the system.
Otherwise the step-(i) is repeated with new values of mean-field parameters   $\left\langle\hat{t}_j\right\rangle=\left\langle\hat{t}^\prime_j\right\rangle$ and $\left\langle\hat{n}_j\right\rangle=\left\langle\hat{n}^\prime_j\right\rangle$.
\end{itemize}
We set the cutoff as $10^{-10}$ and start the simulation with different initial mean-field parameter sets for a fixed set of parameters $J$, $D$, $J_\perp$, $B_z$, and $J_z$.
In general, the simulations with different initial mean field parameter sets give different ground-states at the boundary of two phases and we selected the phase with minimum energy as the ground state.

After obtaining the ground state via self-consistent determination of the mean field parameters, four order-parameters are calculated to identify the nature of the ground state phase of the system.
The magnitude of superfluid order parameter is given by,
\begin{equation}
	|b|=\text{max}\left[|\left\langle\hat{t}_j\right\rangle|\right],
\end{equation}
  where $\text{max}$ denotes maximum value of the parameter obtained out of six-sites within the dashed blue-border in Fig.\,\ref{Cluster}.
Additioally, the average number of particles per site $\left\langle n\right\rangle_{av}$ and difference in number of particles between NN sites $\Delta n$ are also enumerated.
The superfluid order-parameter is a complex quantity and for twisted super-fluid phases in our study the phase difference of superfluid order parameter $b$ among NNN sites is obtained to be $\theta=120^o$ and otherwise $\theta=0^o$.

  \section{\label{secIV} Numerical results}

  
  \begin{figure}[H]
	\centering
		\includegraphics[width=0.47\textwidth]{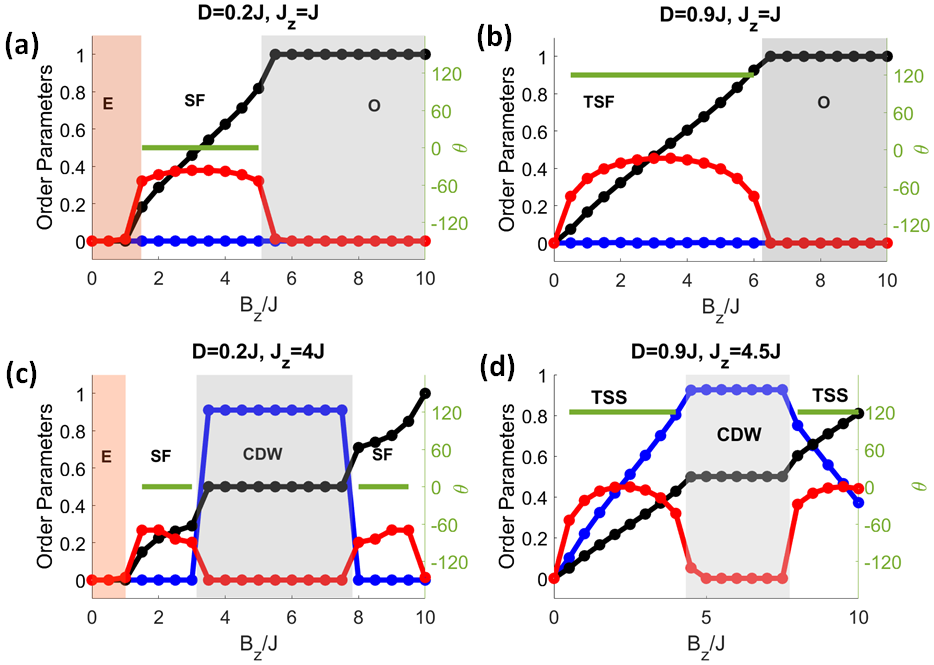}
	\caption{\,(Color online)\,The order parameters are plotted for parameter values (a) $D=0.2J,\, J_z=J$, (b) $D=0.9J,\, J_z=J$, (c) $D=0.2J,\, J_z=4J$, (d) $D=0.9J,\, J_z=4.5J$. $J_\perp$ is fixed at value $10J$. Order-parameters $|b|$, $\left\langle n\right\rangle_{av}$ and $\Delta n$ are plotted as function of magnetic field $B_z$ and denoted by red, black and blue dotted lines respectively. The dots on the lines denote the points in parameter space where the CGMFT is performed and the lines just connect the points. Moreover the phase difference of super-fluid order parameter $\theta$ is shown in the right-side vertical-axis and denoted in green colour. Different phases are indicated by different coloured shades. ``E" and ``O" denote empty and fully-occupied phase of the system respectively. All other phases are described in the main text.}
	\label{OrderParameter}
\end{figure}

Using CGMFT, we determine the order parameters $|b|$, $\left\langle b\right\rangle_{av}$, $\Delta n$ and $\theta$ as a function of magnetic field $B_z$ for different sets of the parameters $(D,J_z)$. The evolution of the order parameters and the resulting field driven phases are shown in Figure fig.\,\ref{OrderParameter} for four illustrative points of the $(D,J_z)$ parameter space.
In fig.\,\ref{OrderParameter}, the DMI increases from the left-column of figures to the right-column of figures, whereas the interaction $J_z$ increases from upper-row of figures towards the lower-row of figures.

For weak DMI\,($D=0.2J$), the field driven phase diagram resembles that of the canonical extended Bose Hubbard model for hard core bosons\,\cite{BoseHubbardModel}\,(see Fig.\,\ref{OrderParameter}(a), (c)). The zero field\,($B_z=0$) ground state corresponds to a singlet phase, or equivalently an empty lattice in the bosonic language. All the order parameters vanish in this limit. This remains true at small values of the applied field reflecting a finite gap to lowest excitations due to the singlet-triplet gap of the local dimers. When the applied field exceeds a critical value, the gap is closed and the ground state acquires a finite density of triplons. These field induced triplons form a superfluid (SF) driven by the NN triplon hopping and is characterised by a finite SF order parameter, $|b|$. The mismatch between the occupancy of the two sublattices\,($\Delta n$) remains zero, reflecting the uniform nature of the SF phase. A vanishing twist angle\,($\theta = 0$) completes the characterization of the phase as a normal superfluid. In the weak interaction limit\,($J_z<2J$) with increasing magnetic field, the density of triplons increases monotonically till full saturation is reached at an upper critical field when each dimer is occupied by a triplon. At saturation, all the order parameters (except average density, $\langle n\rangle_{av}$, of triplons) vanish denoting a lattice fully occupied by triplons. 
In this weak interaction limit, the physics is similar to the non-interacting limit as described in section\,\ref{secII} and so the qualitative feature can be well described using band structure as in figure Fig.\,\ref{Material}(c).

For strong interactions\,($J_z>2J$), an intervening charge density wave (CDW) phase, driven by the strong NN-interaction between triplons, appears in addition to the phases discussed above\,(see Fig.\,\ref{OrderParameter}(c)). With increasing magnetic field, when the density of triplons reaches $\langle n\rangle_{av}=1/2$, the triplons form a staggered CDW pattern where one of the sublattices is fully occupied, while the other remains empty. The potential energy-cost due to nearest neighbor interaction is minimized as there are no nearest neighbor pairs. This is accompanied by a complete quenching of superfluidity, since any hopping of triplons will necessarily involve configurations with energetically costly multiple nearest neighbor pairs. The CDW phase has a finite gap to the addition of any more triplons and the density remains constant at $\langle n\rangle_{av}=1/2$ over a finite rage of applied field. This phase is characterized by a vanishing superfluid order, and a non-zero density mismatch between the two sublattices\,($\Delta n$), reflecting the staggered order.  When the increasing field strength reaches a critical value where the Zeeman energy gain due to increasing magnetization (equivalently, adding more triplons) exceeds the potential energy cost of nearest neighbor repulsion, the density of triplons starts to increase again, resulting in another normal SF phase. Finally, as the field is increased above a saturation value, $B_{sat}$, the ground state enters the fully polarized phase.  

The above argument for the appearance of interaction driven CDW phase at half-filling does not apply for weak to moderate interaction strengths\,($J_z < 2J$), as the kinetic energy gain due to the delocalization of triplons exceeds the potential energy cost of NN-interactions. 
  
The sequence of field-driven phase changes markedly for strong DMI. As shown earlier in Sec.\,\ref{secII}, in the non-interacting limit, the triplon band minimum shifts from the $\Gamma$ point to the $K$ and $K'$\,(Fig.\,\ref{Material}(d)) and the BEC of triplons occur at finite momentum. A local minimum persists at the center of the Brillouin zone, and the energy gap between the triplon-sector and the singlet dimer sector, $E_{K}$ in Eq.\,\ref{EnergyGap} decreases with increasing DMI. This behavior persists in the presence of weak to moderate interaction\,($J_z < 2J$) and is reflected in Fig.\,\ref{OrderParameter}(b). For the present choice of parameters, the energy of the lowest triplon excitation is vanishingly small. The triplon density acquires a  finite value for an infinitesimally small $B_z$, and increases monotonically with the strength of the applied field. In this regime, the triplons form a superfluid\,($|b|>0$). More interestingly, the complex NNN hopping process imparts a complex phase to the superfluid order parameter, as seen by a finite expectation value of the twist angle\,($\theta\neq 0$). In other words, the ground state in this parameter range is a twisted superfluid (TSF).  
The triplon density increases monotonically, with the ground state remaining a TSF, till the fully polarized phase is reached at a saturation field, $B_{sat}$.

Finally in the strong DMI and strong interaction limit\,(see Fig.\,\ref{OrderParameter}(d)),the twisted superfluid is replaced by a twisted supersolid phase, in addition to the appearance of an interaction-driven CDW phase at  $\langle n\rangle_{av} = 1/2$ over a finite range of applied field. In the twisted supersolid phase (TSS), the ground state is characterized by a finite $\Delta n$ (density mismatch between the two sublattices), in addition to a complex superfluid order parameter\,($|b|\neq 0,\, \theta\neq 0,$). The finite density difference between two sublattices provide the diagonal order concurrently with the finite (twisted) superfluid ordering. It is surprising that the ground state exhibits TSS order even at low triplon densities. This is understood by recalling that the primary delocalization process in this parameter regime involves the DMI-induced intra-sublattice complex next-nearest-neighbor hopping. The strong NN-repulsion between the triplons further suppresses inter-sublattice hopping processes, resulting in a preferential occupation of one of the two sublattices at small densities.

  \begin{figure}[H]
	\centering
		\includegraphics[width=0.47\textwidth]{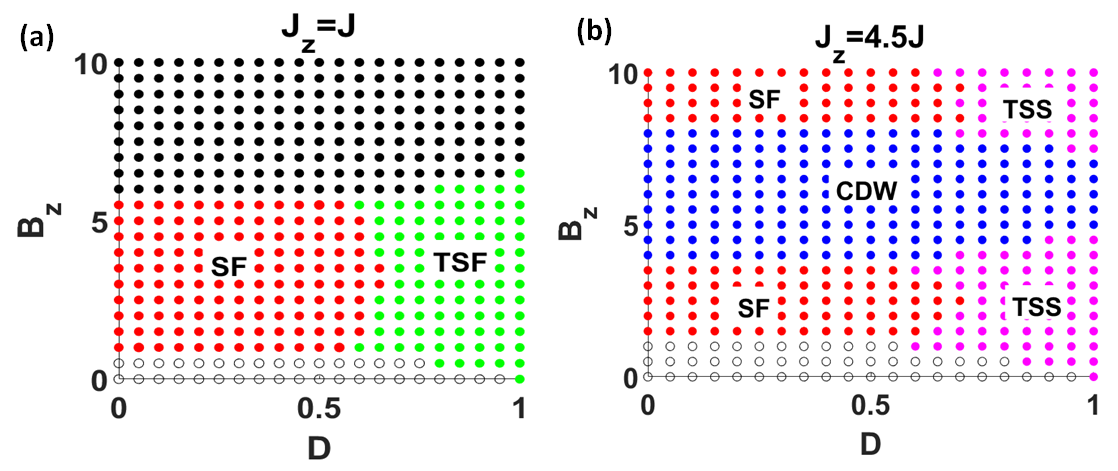}
	\caption{\,(Color online)\,Phase diagram at two different interaction values (a) $J_z=J$ and (b) $J_z=4.5J$. Each dot or circle denotes the parameter point where the CGMFT is performed. Different color denotes different phases in parameter space. The empty and dotted black circles represent the empty and fully occupied lattice phases, respectively.}
	\label{PhaseDiagram}
\end{figure}
  
The phase diagram in $B_z-D$ parameter space is shown in the figure Fig.\,\ref{PhaseDiagram} for two different values of interaction $J_z=J$ and $J_z=4.5J$.
A comparison of the phase diagrams at moderate\,($J_z<2J$) and strong\,($J_z>2J$) interactions reveal,
  \begin{itemize}
  \item[(i)] Appearance of CDW phase at half-filling in the strong interaction limit. The CDW phase appears at half filling dividing the SF-region which appear at weak to moderate $J_z$ into two SF-regions.
  \item[(ii)] In the strong interaction limit the TSF phase is replaced by a TSS phase.
\end{itemize}

\section{\label{secV}Material Realization}

There are two main ingredients to realize TSF or TSS phases; firstly, a bilayer honeycomb valence bond state is required; secondly, a DMI greater than a critical value is required. Although, to the best of our knowledge, there are no materials described in the literature which satisfies our model,
a rough idea is sketched here to obtain TSS or TSF phases based on the real materials \ce{Bi3Mn4O12(NO3)} and \ce{CrI3}.

The material \ce{Bi3Mn4O12(NO3)} consists of Mn$^{4+}$ ions carrying spin $S=\frac{3}{2}$ arranged in a bilayer honeycomb lattice with A-A stacking. No magnetic ordering is observed down to the lowest temperatures. Theoretical studies indicate that the disordered magnetic state may be an interlayer dimer phase which is adiabatically connected to the direct products of singlets\,\cite{Important,InterLayer1,InterLayer2,InterLayer3,InterLayer4}, although a spin liquid phase cannot be completely ruled out\,\cite{Ref1,Ref2,Ref3,Ref4,Ref5,Ganesh,Ref7,Ref8,Ref9,Ref10,Ref11,Ref12}. If the ground state phase is a spin liquid, an interlayer dimer phase can be induced by applying a pressure along the perpendicular direction~\cite{Ganesh,InterLayer3}.

On the other hand, experimentally well studied van der Waals material \ce{CrI3} is a honeycomb ferromagnet where Cr$^{2+}$-ions carry spin-$\frac{1}{2}$ momentum forming a honeycomb lattice.
This material has been well studied by tuning the number of layers\,\cite{CrI3_Layer} as well as varying the pressure\,\cite{CrI3_Pressure}.
It is shown that the interlayer antiferromagnetic Heisenberg exchange interaction changes linearly with the distance between two layers by application of pressure on the material and can be achieved a interlayer coupling twice as compared with the initial interlayer coupling\,\cite{CrI3_Pressure}.
Although a valence bond state is not detected in \ce{CrI3} (and there are indications of structural phase transition under pressure)\,\cite{CrI3_Pressure}, the pressure induced high interlayer coupling in \ce{CrI3} motivates for searching of pressure induced valance bond states in a family of Van der Walls honeycomb magnets \ce{CrBr3}\,\cite{CrBr3}, \ce{CrGeTe3}\,\cite{CrGeTe3}, \ce{CrSiTe3}\,\cite{CrSiTe3}, \ce{FePS3}\,\cite{FePS3}, \ce{NiPS3}\,\cite{NiPS3}.

Another challenge for achieving the TSF and TSS phases is to obtain a high DMI\,($|D|>J/\sqrt{3}$).
The materials \ce{Bi3Mn4O12(NO3)} and \ce{CrI3} are experimentally predicted to possess DMI\,\cite{RealMaterial1,RealMaterial2}.
However the magnitude of DMI is small for realizations TSS and TSF states.
This problem can be overcome by application of circularly polarized light.
A circularly polarized light couples either to the charge\,\cite{Optical1} or to the magnetic degrees of freedom\,\cite{Optical2} and it has been shown theoretically that both kinds of mechanisms give rise to scalar spin chiral interaction in a honeycomb magnetic insulator.
The synthetic scalar spin chiral interaction as in reference Ref.\,\cite{Optical1} given by,
\begin{align}
    \pazocal{H}_{\chi}&=\chi\sum_{\left\langle\left\langle j,k\right\rangle\right\rangle,m}
    \nu_{jk}\hat{S}_{i,m}\cdot\left(\hat{S}_{j,m}\times\hat{S}_{k,m}\right)
    \nonumber\\
    &\approx
    \frac{3\mathfrak{i}\chi}{4} \sum_{\left\langle\left\langle j,k\right\rangle\right\rangle}
    \nu_{jk} \left\langle \hat{n}_i\right\rangle 
    \left[
    \hat{t}_j^\dagger\hat{t}_k-\text{H.c.}
    \right],
    \label{Eq10}
\end{align}
where the $i$-th site is the NNN neighbor of both $j$-th and $k$-th sites, $\chi$ is synthetic scalar spin chirality.
Comparing the equations Eq.\,\ref{Hamiltonian} and Eq.\,\ref{Eq10}, the effective DMI for bond $\left\langle\left\langle jk\right\rangle\right\rangle$ due to circularly polarized light is $D=\frac{3\chi\left\langle n_i\right\rangle}{2}$.
The scalar spin chirality $\chi$ is shown to have a resonance for the frequency of light near $\omega_n=\frac{1}{\hbar}\frac{U}{n}$, where $U$ is onsite interaction representing electron-electron repulsion and $n$ is a positive integer.
Thus the effective DMI can be tuned as high as possible tuning the frequency of the light nearby $\omega_n$.
However the results in Ref.\,\cite{Optical1} is based on single-band extended Hubbard model; for a magnetic material which cannot be described by a one band Hubbard model, one needs a more careful theoretical treatment.
On the other hand, application of light may not be required for a real material if the cutoff of DMI to realize TSF or TSS states is reduced due to presence of frustration among interlayer and intralayer interactions\,(see Appendix.\,\ref{appendixC}).

\section{\label{secVI}Conclusion}

To summarize, we have shown that a magnetic analog of the novel twisted superfluid (TSF) state reported in recent experiments with ultracold atoms in an optical lattice\,\cite{Rb} can be realized in a bilayer quantum antiferromagnet with realistic interactions. We show that TSF phase is induced by DMI greater than a critical value. For Ising-like anisotropy of the intra-plane Heisenberg interactions, the TSF phase is replaced by a twisted supersolid (TSS) phase. While the strength of DMI required for the stabilization of TSF and TSS phases\,($D/J \gtrsim 0.5$) is not observed natively in most quantum magnets, recent experiments have shown that a strong DMI can be induced in thin films of insulating magnets, by forming heterostructures with heavy metals\,(with strong spin-orbit coupling)\,\cite{heavymetal1,heavymetal2,heavymetal3,heavymetal4}. 
Our results show that circularly polarized light can also induce large DMI by varying the frequency of light\,\cite{Optical1}. Finally, the presence of frustration among interlayer and intralayer Heisenberg exchange interactions in a material can lower the value of the critical DMI required to realize TSF and TSS phases\,(Appendix.\,\ref{appendixC}), thus facilitating their experimental observation.
We propose that the material \ce{Bi3Mn4O12(NO3)}\,\cite{Important,InterLayer1,InterLayer2,InterLayer3,InterLayer4} and family of honeycomb magnets \ce{CrBr3}\,\cite{CrBr3}, \ce{CrGeTe3}\,\cite{CrGeTe3}, \ce{CrSiTe3}\,\cite{CrSiTe3}, \ce{FePS3}\,\cite{FePS3}, \ce{NiPS3}\,\cite{NiPS3} are promising candidate materials to realize TSF and TSS phases. 

\section*{Acknowledgement}
It is a pleasure to thank Oleg Sushkov for useful discussions. Financial support from the Ministry of Education, Singapore, in the form of grant MOE2018-T1-001-021 is gratefully acknowledged.
A. Chatterjee acknowledges the financial support received from
INSPIRE, Department of Science and Technology, Govt. of India and the NTU-India Connect Program.
A. Chatterjee would like to thank the School of Physical and Mathematical Sciences at NTU for their kind hospitality.

\appendix

\section{Bond-operator formalism}
\label{appendixA}
The spin-operators in terms of bond-operators are given by\,\cite{bond_operator2},
\begin{align}
    \hat{S}_{j,l}^+=\frac{\hat{t}_{1,j}^\dagger \hat{t}_{0,j} + \hat{t}_{0,j}^\dagger \hat{t}_{\bar{1},j}}{\sqrt{2}}\pm
    \frac{\hat{s}_j^\dagger \hat{t}_{\bar{1}j}-\hat{t}_{1,j}^\dagger \hat{s}_j}{\sqrt{2}}\nonumber\\
    \hat{S}_{j,l}^-=\frac{\hat{t}_{\bar{1},j}^\dagger \hat{t}_{0,j} + \hat{t}_{0,j}^\dagger \hat{t}_{1,j}}{\sqrt{2}}\mp
    \frac{\hat{s}_j^\dagger \hat{t}_{1,j}-\hat{t}_{\bar{1},j}^\dagger \hat{s}_j}{\sqrt{2}}\nonumber\\
    \hat{S}_{j,l}^z=\frac{\hat{t}_{1,j}^\dagger \hat{t}_{1,j} - \hat{t}_{\bar{1},j}^\dagger \hat{t}_{\bar{1},j}}{2}\pm
    \frac{\hat{s}_j^\dagger \hat{t}_{0,j}+\hat{t}_{0,j}^\dagger \hat{s}_j}{2},
    \label{eq::SpinTriplon}
\end{align}
where the upper-sign is for layer $l=$A and the lower sign is for the layer $l=$B. $\hat{t}_{1,j}^\dagger$, $\hat{t}_{\bar{1},j}^\dagger$ and $\hat{t}_{0,j}^\dagger$ create states $\ket{\uparrow\uparrow}$, $\ket{\downarrow\downarrow}$ and $(\ket{\uparrow\downarrow}+\ket{\downarrow\uparrow})/\sqrt{2}$ on $j$-th bond respectively.
After the bond operator transformation, the bilayer honeycomb spin system transforms into an effective single layer honeycomb lattice system with triplon and singlet operators on each site. 
At low temperature, in the limit $|J_\perp|\gg |J|$, it can be assumed that the ground state is product of singlets on the interlayer NN-bonds. 
Assuming singlets form the empty or vacuum state and taking advantage of the hardcore nature of the bosons, we can simply transform the following quadratic operators into a single triplon operators in equation Eq.\,\ref{eq::SpinTriplon},
\begin{equation*}
    \hat{s}_j^{\dagger} \hat{t}_{\alpha j}\rightarrow \hat{t}_{\alpha j},\,\,
    \hat{t}_{\alpha j}^\dagger \hat{s}_j \rightarrow \hat{t}_{\alpha j}^\dagger .
\end{equation*}
Moreover application of magnetic field in the $z$-direction lowers the energy of triplon correspond to the operator $\hat{t}^\dagger_{1,j}$ and so neglecting all other triplon operators in equation Eq.\,\ref{eq::SpinTriplon}, we get the following bond operator transformation\,\cite{ZapfBatista},
\begin{equation*}
    \hat{S}_{j,l}^+=\mp\frac{1}{\sqrt{2}}\hat{t}_j^\dagger,\,
    \hat{S}_{j,l}^-=\mp \frac{1}{\sqrt{2}} \hat{t}_{j},\,
    \hat{S}_{j,l}^z= \frac{1}{2} \hat{t}_{j}^\dagger\hat{t}_{j},
\end{equation*}
where the subscript-$1$ is omitted from the operator $\hat{t}_{1,j}^\dagger$.

\section{Interaction term in reciprocal space}
\label{appendixB}
The last term in equation Eq.\,\ref{Hamiltonian} represents the NN repulsive interaction and in reciprocal space it is given by,
\begin{equation}
    \pazocal{H}_V=\frac{J_z}{2}
    \sum_{\boldsymbol{\alpha}}
    \sum_{\bold{k},\bold{k}^\prime,\bold{k}^{\prime\prime}}
    e^{i(\bk-\bk^\prime)\cdot\boldsymbol{\alpha}}\,\,
    \hat{t}^\dagger_{\bk}
    \hat{t}_{\bk^\prime}
    \hat{t}^\dagger_{\bk^{\prime\prime}}
    \hat{t}_{\bk+\bk^{\prime\prime}-\bk^\prime}.
\end{equation}
Considering triplons are only present at $K$  and $K'$ points as well as neglecting quantum fluctuations around these points, the interaction Hamiltonian can be explicitly written as,
\begin{widetext}

\begin{align}
    \pazocal{H}_{V}&=
    \frac{3J_z}{2}\left[
    \hat{t}^\dagger_{\bold{K}}
    \hat{t}_{\bold{K}}
    \hat{t}^\dagger_{\bold{K}}
    \hat{t}_{\bold{K}}
    +
    \hat{t}^\dagger_{\bold{K}^\prime}
    \hat{t}_{\bold{K}^\prime}
    \hat{t}^\dagger_{\bold{K}^\prime}
    \hat{t}_{\bold{K}^\prime}
    +
    \hat{t}^\dagger_{\bold{K}}
    \hat{t}_{\bold{K}}
    \hat{t}^\dagger_{\bold{K}^\prime}
    \hat{t}_{\bold{K}^\prime}
    +
    \hat{t}^\dagger_{\bold{K}^\prime}
    \hat{t}_{\bold{K}^\prime}
    \hat{t}^\dagger_{\bold{K}}
    \hat{t}_{\bold{K}}
    \right]
    \nonumber\\
    &+
    \frac{J_z}{2}
    \sum_{\boldsymbol{\alpha}}
    \left[
    e^{i(\bold{K}-\bold{K}^\prime)\cdot\boldsymbol{\alpha}}
    \hat{t}^\dagger_{\bold{K}}
    \hat{t}_{\bold{K}^\prime}
    \hat{t}^\dagger_{\bold{K}^\prime}
    \hat{t}_{\bold{K}}
    +
    e^{i(\bold{K}^\prime-\bold{K})\cdot\boldsymbol{\alpha}}
    \hat{t}^\dagger_{\bold{K}^\prime}
    \hat{t}_{\bold{K}}
    \hat{t}^\dagger_{\bold{K}}
    \hat{t}_{\bold{K}^\prime}
    \right]
    \nonumber\\
    &=\frac{3J_z}{2} 
    \left[
    \hat{n}_{\bold{K}}^2
    +
    \hat{n}_{\bold{K}^\prime}^2
    +
    2\hat{n}_{\bold{K}}\hat{n}_{\bold{K}^\prime}
    \right]
    \nonumber\\
    &=\frac{3J_z}{2}\left(
    \hat{n}_{\bold{K}}
    +
    \hat{n}_{\bold{K}^\prime}
    \right)^2,
\end{align}

\end{widetext}
thus coexistence of particles at $\bold{K}$ and $\bold{K}^\prime$ points does not seem to increase the energy of the system.
However, quantum fluctuations around the points $\bold{K}$ and $\bold{K}^\prime$ increases the energy due to presence of triplons at both $\bold{K}$ and $\bold{K}^\prime$ points as shown in the reference Ref.\,\cite{SymmetryBreaking1}.

\section{Introducing more terms in the spin Hamiltonian}
\label{appendixC}
\begin{figure}[H]
	\centering
		\includegraphics[width=0.47\textwidth]{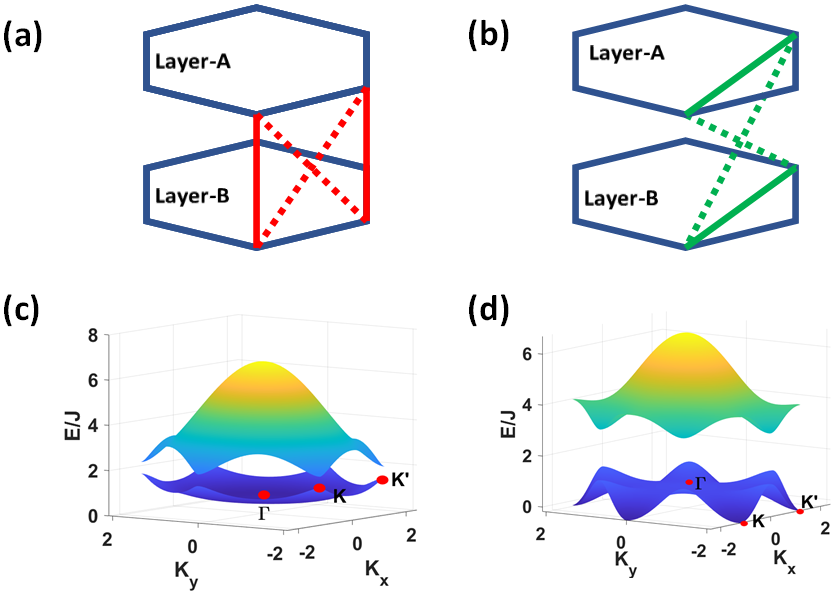}
	\caption{\,(Color online)\, (a) Intradimer and interlayer NN bonds are indicated by red solid and red dotted lines, respectively. (b) The intralayer and interlayer NNN bonds are indicated by green solid and green dotted lines, respectively. (c) Triplon band structure at $D=0.1J$, (b) Triplon band structure at $D=0.8J$. The other parameters for the band structures are $J_\perp=10J$, $J_1=0.8J$, $J_{2\perp}=3J$, $B_z=0.0$, $J_z=0.0$, $J_{3\perp}=0$.}
	\label{fig::BandAppendix}
\end{figure}

In this appendix, we have taken into account additional interlayer bonds (see Fig.\,\ref{fig::BandAppendix}(a), (b)). We introduce intralayer NNN Heisenberg-exchange interaction $J_1$ as well as interlayer NN Heisenberg exchange interaction $J_{2\perp}$ in the spin Hamiltonian in equation Eq.\,\ref{eq::SpinHamiltonian},
\begin{widetext}
\begin{align}
\pazocal{H}=&J_\perp\sum_{\substack{i,\,m\in A\\ n\in B}}\bold{S}_{i,m}\cdot\bold{S}_{i,n}-B_z\sum_{i,m} S^z_{i,m}
+\sum_{\left\langle i,j\right\rangle,m}\left[J(\bold{S}^x_{i,m}\bold{S}^x_{j,m}+\bold{S}^y_{i,m}\bold{S}^y_{j,m})+J_z\bold{S}^z_{i,m}\bold{S}^z_{j,m}\right] 
\nonumber\\
&+ D\sum_{\left\langle\left\langle i,j\right\rangle\right\rangle,m} \nu_{ij} \hat{z}\cdot (\bold{S}_{i,m}\times \bold{S}_{j,m})
+ D_\perp\sum_{\substack{\left\langle\left\langle i,j\right\rangle\right\rangle\\ m\in A,B,\,n\neq m}} \nu_{ij} \hat{z}\cdot (\bold{S}_{i,m}\times \bold{S}_{j,n})
\nonumber\\
&+J_1\sum_{\left\langle\left\langle i,j\right\rangle\right\rangle,m} 
\bold{S}_{i,m}\cdot\bold{S}_{i,m}
+J_{2,\perp}\sum_{\substack{\left\langle i,j\right\rangle\\m\in A,B,\,n\neq m}}
\bold{S}_{i,m}\cdot\bold{S}_{i,n}
+J_{3,\perp}\sum_{\substack{\left\langle\left\langle i,j\right\rangle\right\rangle\\m\in A,B,\,n\neq m}}
\bold{S}_{i,m}\cdot\bold{S}_{i,n}.
\end{align}
Moreover we have added symmetry allowed interlayer NNN DMI $D_\perp$ and Heisenberg interaction $J_{3\perp}$.
The DMI on dimer-bond and NN interlayer bonds are zero due to presence of inversion center at the middle of the bonds.
The corresponding real space triplon Hamiltonian is given as,
\begin{align}
    \pazocal{H}=&\frac{J-J_{2\perp}}{2}\sum_{\left\langle ij\right\rangle} \left[\hat{t}^\dagger_i\hat{t}_j+\text{H.c.}\right]
    +\frac{J_1-J_{3\perp}}{4}\sum_{\left\langle\left\langle ij\right\rangle\right\rangle} \left[\hat{t}^\dagger_i\hat{t}_j+\text{H.c.}\right]
+\frac{\mathfrak{i}(D+D_\perp)}{2}\sum_{\left\langle\left\langle ij\right\rangle\right\rangle} \nu_{ij} \left[\hat{t}^\dagger_i\hat{t}_j-\text{H.c.}\right]
\nonumber\\
&+\left(\frac{J_\perp}{4}-B_z\right)\sum_i \hat{t}^\dagger_i \hat{t}_i
+\frac{J_{2\perp}+J_z}{2}\sum_{\left\langle ij\right\rangle} \hat{n}_i \hat{n}_j
+\frac{J_1+J_{3\perp}}{4}\sum_{\left\langle\left\langle ij\right\rangle\right\rangle} \hat{n}_i \hat{n}_j
\end{align}
\end{widetext}
It is noticeable that the Heisenberg interaction $J_{2\perp}$ renormalizes the interactions $J$ and $J_z$, whereas $J_1$ and $J_{3\perp}$ add additional NNN hopping term and NNN interaction term. 
Moreover $D_\perp$ renormalizes $D$ by simply adding up and so $D_\perp$ is absorbed into $D$ in the rest of this section.
In this appendix, we investigate the triplon Hamiltonian, neglecting the interaction terms.
The non-interacting Hamiltonian in $k$-space is similar to equation Eq.\,\ref{eq::Hkequation} and given by,
\begin{equation}
 \pazocal{H}_0=\sum_{\bold{k}} \Psi_{\bold{k}}^\dagger  \left[g'(\bold{k})\sigma_0+\bold{h'}\cdot\boldsymbol{\sigma}\right]\Psi_{\bold{k}},
\end{equation}
where,
\begin{align*}
    g'(\bold{k})&=(J_\perp/4)-B_z+((J_1-J_{3\perp})/2)\sum_i\cos({\bold{k}\cdot\boldsymbol{\beta}_i}),\\
    h'_x(\bold{k})&=((J-J_{2\perp})/2)\sum_{i} \cos(\bold{k}\cdot\boldsymbol{\alpha}_i),\\ 
    h'_y(\bold{k})&=((J-J_{2\perp})/2)\sum_{i} \sin(\bold{k}\cdot\boldsymbol{\alpha}_i),\\
    h'_z(\bold{k})&=D\sum_{i} \sin(\bold{k}\cdot\boldsymbol{\beta}_i)
\end{align*}
, where $\boldsymbol{\alpha}_i$ and $\boldsymbol{\beta}_i$ are the NN and NNN vectors for each layer respectively.
The band structure for two different $D$-values are plotted in the figure Fig.\,\ref{fig::BandAppendix}(c) and (d).
The minima of the bands are at the $\Gamma$-point and $K$ (or $K'$)-point for $D=0.1J$ and $D=0.8J$ respectively.
The results are same when $J_{2\perp}=0$ and $J_1=0$ as in the figures Fig.\,\ref{Material}(c) and (d).
The condition for the band minima at $K$ or $K'$-point is given as,
\begin{equation}
    |D|>\frac{2|J-J_{2\perp}|-3(J_1-J_{3\perp})}{2\sqrt{3}}.
\end{equation}
We note that the critical DMI required to realise TSF or TSS phases depends on various Heisenberg exchange interactions.
The conditions $J\approx J_{2\perp}$ and $J_1\approx J_{3\perp}$ lowers the critical value of the DMI to realize TSF or TSS phases.
Thus we can conclude that frustration among interlayer and intralayer Heisenberg interactions can lower the critical DMI required to realize TSF and TSS phases.
\\
\\

\bibliographystyle{apsrev4-1}

%

\end{document}